# DOMESTIC ACTIVITIES CLUSTERING FROM AUDIO RECORDINGS USING CONVOLUTIONAL CAPSULE AUTOENCODER NETWORK

*Ziheng Lin, Yanxiong Li\*, Zhangjin Huang, Wenhao Zhang, Yufeng Tan, Yichun Chen, Qianhua He*

School of Electronic and Information Engineering, South China University of Technology, Guangzhou, China
eeyxli@scut.edu.cn

**ABSTRACT**

Recent efforts have been made on domestic activities classification from audio recordings, especially the works submitted to the challenge of DCASE (Detection and Classification of Acoustic Scenes and Events) since 2018. In contrast, few studies were done on domestic activities clustering, which is a newly emerging problem. Domestic activities clustering from audio recordings aims at merging audio clips which belong to the same class of domestic activity into a single cluster. Domestic activities clustering is an effective way for unsupervised estimation of daily activities performed in home environment. In this study, we propose a method for domestic activities clustering using a convolutional capsule autoencoder network (CCAN). In the method, the deep embeddings are learned by the autoencoder in the CCAN, while the deep embeddings which belong to the same class of domestic activities are merged into a single cluster by a clustering layer in the CCAN. Evaluated on a public dataset adopted in DCASE-2018 Task 5, the results show that the proposed method outperforms state-of-the-art methods in terms of the metrics of clustering accuracy and normalized mutual information.

***Index Terms***— Domestic activity clustering, convolutional capsule autoencoder network, human activity estimation

## 1. INTRODUCTION

There is a rising interest in home environments that enhance the living quality for humans in terms of safety, comfort, and home care [1]. Domestic activities classification from audio recordings were investigated in previous works [1], [2], whose task is to identify acoustic scenes using home environmental sounds [3]. From a technical point of view, it can be regarded as a task of acoustic scene classification.

Many works were done on acoustic scenes classification in recent years [2]-[10]. In these efforts, it was assumed that the identities of acoustic scenes were known in advance, and at least one classifier was first trained using manual labels of audio recordings. Accordingly, the main aim of these efforts was to determine the predefined class of the acoustic scene to which each test audio recording belongs. However, the identities of classes of acoustic scenes are not always available in practice due to the following reasons: label ambiguity, label loss, weak or incorrect labels, and high cost for manually labeling massive audio recordings [11]. When we don't have any labels or prior information concerning massive audio recordings of acoustic scenes, the initial tasks could be how to merge audio recordings which belong to the same class of acoustic scene into a single cluster instead of determining their specific identities of acoustic scenes. In this scenario, the problem for domestic activities estimation in home environment becomes a problem of domestic activities clustering. To the best of our knowledge, there is no prior work on domestic activities clustering from audio recordings up to date.

In addition, the common hand-crafted features used for acoustic scene classification (or clustering) include the logarithm mel-band energy, mel frequency cepstral coefficient (MFCC), spectral flux, spectrogram, Gabor filterbank, cochleograms, I-vector, histogram of gradients features [12]-[15], the histogram of gradients of time-frequency representations (HGTR) [14], hash features [16], and local binary patterns [17], [18]. In recent years, some transformed features using matrix factorization [19], [20] and deep neural network [6], [11], [21], are used to address the lack of flexibility of hand-crafted features. Hand-crafted or shallow features did not effectively represent the property differences among various classes of acoustic scenes, and thus their performance was inferior to that of deep transformed features learned by deep neural networks, such as convolutional neural network (CNN) [11], [22]-[25].

Although the CNN has advantages for learning deep embedding with better performance, it has its own deficiencies for feature learning. First, the smallest unit in the output of convolution layers is a scalar, which is able to identify whether a feature element exists or not, but unable to effectively represent the connections between feature elements. Second, the usage of the pooling layer leads to the loss of valuable information, such as the relative positions of feature elements. As a result, the CNN is unable to effectively recognize the spatial relationships before and after the rotation of an object and the spatial relationships between objects [26]. In contrast, the capsule network [26], can efficiently learn the relative position relationship between several feature elements, and make predictions with higher accuracy than CNN when angle changes. For example, Amiriparian et al. recently adopted a capsule network to successfully deal with the spatial hierarchy between images that were extracted from audio files [27].

---

\* Corresponding author: Yanxiong Li (eeyxli@scut.edu.cn).
This work was partly supported by the NSFC (61771200), the joint NSFC-AF project (Sound event detection for complex audio with noisy labels), Guangdong basic and applied basic research foundation in 2021 (Research on speaker segmentation and clustering for massive conversational speech), and the national undergraduate training program for innovation and entrepreneurship (201910561024).

Inspired by the success of CNN and capsule network for feature representation learning, we propose a method for domestic activity clustering from audio recordings using a convolutional capsule autoencoder network (CCAN). The CCAN is a combination of CNN and capsule network, which owns the advantages of CNN and is also able to overcome the CNN's deficiencies for feature representation learning. In the proposed method, the deep embeddings are learned by an autoencoder in the CCAN, and merged into different clusters by a clustering layer in the CCAN. The performance of the proposed method is evaluated on a public dataset adopted in DCASE-2018 Task 5. Therefore, the main contribution of this study is to address a new problem of domestic activity clustering from audio recordings using a CCAN.

The rest of the paper is organized as follows. Section 2 describes the proposed method, and Section 3 presents the experiments. Finally, conclusions and future works are given in Section 4.

## 2. METHOD

The framework of the proposed method for domestic activity clustering is illustrated in Fig. 1, which is mainly composed of three modules: an Encoder, a Decoder and a clustering layer. The autoencoder consists of the Encoder and the Decoder and is used for learning deep embedding, while the clustering layer is used to merge deep embeddings into various clusters. A joint loss is used to simultaneously guide deep embedding learning in the autoencoder and deep embedding mergence in the clustering layer. As shown in Fig. 1, the main difference between the proposed method and previous methods [22]-[25] is that we introduce the layers of primary capsules and class capsules into the convolutional autoencoder for deep embedding learning.

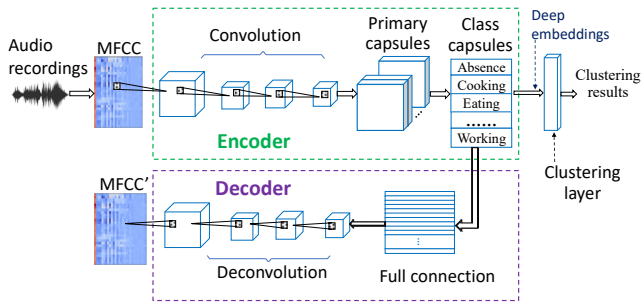

Fig. 1. The framework of the proposed method for domestic activities clustering from audio recordings.

### 2.1. CCAN structure

MFCC is extracted from each audio clip, since it is one of the most effective acoustic features for acoustic scene classification [12]. In the Encoder, MFCCs are fed into convolutional layers which are connected sequentially for obtaining transformed feature vectors, and then the layers of primary capsules and class capsules are introduced to extract deep embedding based on their inputs. The 2-norms of the vectors from class capsule layer are used as deep embeddings and fed into clustering layer. The vectors from class capsule layer are reshaped as a supervector by a full-connection layer in the Decoder. The supervector is fed into deconvolutional layers that are used for reconstructing their input features, i.e., for finally obtaining MFCC′.

In the layers of primary capsules and class capsules, the input prediction vectors $\hat{u}_{l|k}$ are defined by

$$\hat{u}_{l|k} = W_{kl} u_k, \quad (1)$$

where $W_{kl}$ is a weight matrix between capsule $k$ in current layer and capsule $l$ in next layer; and $u_k$ denote the output of capsule $k$ in previous layer. The number of rows and the number of columns in $W_{kl}$ are determined by the vector length of primary capsule and class capsule, respectively.

A coupling coefficient $c_{kl}$ is defined by

$$c_{kl} = \frac{\exp(b_{kl})}{\sum_m \exp(b_{km})}, \quad (2)$$

where $b_{kl}$ is the weight between capsule $k$ in current layer and capsule $l$ in next layer. A squash function $v_l$ is used to ensure that the minimum length of the output vector is 0 and the maximum length does not exceed 1, and is defined by

$$v_l = squash(s_l) = \frac{\|s_l\|_2^2 \, s_l}{(1+\|s_l\|_2)(1+\|s_l\|_2)}. \quad (3)$$

where $s_l$, an intermediate vector in the capsule $l$, is the direct result of dynamic routing; and $\|s_l\|_2$ stands for the 2-norm of $s_l$. In summary, the dynamic routing algorithm between two adjacent capsules layers is presented in Table 1.

Table 1. Dynamic routing algorithm between capsules layer $n$ and capsules layer $n+1$.

| |
|---|
| **Input:** |
|    Prediction vectors $\hat{u}_{l|k}$ |
| **Do:** |
|    $b_{kl} = 0$ |
|   **Repeat** *max_iteration* times: |
|     $c_{kl} = softmax(b_{kl})$ |
|     $s_l = \sum_k c_{kl} \, \hat{u}_{l|k}$ |
|     $v_l = squash(s_l)$ |
|     $b_{kl} = b_{kl} + \hat{u}_{l|k} \, v_l$ |
| **End** |
| **Output:** |
|    Capsules $v_l$ in layer $n+1$ |

## 2.2. Joint loss

The proposed framework is composed of an Encoder $f_E(\cdot)$, a Decoder $f_D(\cdot)$, and a clustering layer. One of its aims is to find a code (deep embedding here) for each input feature by minimizing the mean squared errors between its input and output over all input features. The reconstruction loss is defined by

$$L_r = \min \frac{1}{N}\sum_{n=1}^{N}\|f_D(f_E(x_n)) - x_n\|_2^2. \quad (4)$$

where $f_E(x)$ and $f_D(h)$ are defined by

$$f_E(x) = \sigma(x * W), \quad (5)$$
$$f_D(h) = \sigma(h * W'). \quad (6)$$

where $x$ and $h$ denote input matrices (feature vectors) of the Encoder and the Decoder, respectively; $W$ and $W'$ are weight matrices of the Encoder and the Decoder, respectively; $*$ represents convolution operator; and $\sigma(\cdot)$ stands for an activation function, such as rectified linear unit or sigmoid.

The clustering layer of the CCAN is introduced from [25], which maintains cluster centers $\{u_j\}_1^K$ as trainable weights and maps each class capsule $z_i$ into a soft label $q_i$ by Student's $t$-distribution [28]:

$$q_{ij} = \frac{(1+\|z_i - u_j\|^2)^{-1}}{\sum_j (1+\|z_i - u_j\|^2)^{-1}}, \quad (7)$$

where $q_{ij}$ denotes the predicted probability that $z_i$ belongs to cluster $j$.

The clustering loss $L_c$ is defined as a Kullback-Leibler (KL) divergence [29] between the distribution of soft labels $Q = \{q_i\}$ and the predefined target distribution $P = \{p_i\}$. $L_c$ is calculated by

$$L_c = KL(P \| Q) = \sum_i \sum_j p_{ij} \log \frac{p_{ij}}{q_{ij}}. \quad (8)$$

where $p_{ij}$ denotes the predefined target (i.e., ground-truth) probability that $z_i$ belongs to cluster $j$. As done in [24], $p_{ij}$ is defined by

$$p_{ij} = \frac{q_{ij}^2 / \sum_i q_{ij}}{\sum_j (q_{ij}^2 / \sum_i q_{ij})}, \quad (9)$$

A joint loss $L_J$ is designed for simultaneously guiding deep embeddings learning by autoencoder in the CCAN and deep embeddings clustering by clustering layer in the CCAN. It is defined by

$$L_J = L_r + \alpha L_c, \quad (10)$$

where $\alpha > 0$, is a balance coefficient for controlling the contributions of $L_r$ and $L_c$ to the value of joint loss $L_J$. The optimization of the CCAN guided by joint loss $L_J$ is presented in Table 2. After the update of the CCAN guided by $L_J$, the optimized clustering results are obtained.

Table 2. The optimization of the CCAN guided by $L_J$.

| |
|---|
| **Initialization:** |
|     Pretrain autoencoder by setting $\alpha = 0$ (i.e., $L_J = L_r$) for obtaining target distribution; |
|     Initialize cluster centers by $K$-means clustering on deep embeddings learned by autoencoder; |
|     Set $\alpha$ to be a fixed non-zero value. |
| **Repeat:** |
|     Update autoencoder's weights and cluster centers using backpropagation and mini-batch stochastic gradient descent algorithms [30]; |
|     Update the predicted probability $q_{ij}$ by Eq. (7) and target probability $p_{ij}$ by Eq. (9). |
| **Until** the change of label assignments between two adjacent iterations is less than a threshold $\varepsilon$. |
| **Output:** |
|     Optimized clustering results. |

## 3. EXPERIMENTS

This section first introduces experimental dataset and setup, and then presents experimental results and discussions.

### 3.1. Experimental dataset

Experimental dataset adopted in this study is a publicly available dataset adopted in DCASE-2018 Task 5, which is a derivative of the SINS dataset [1] collected by a network of 13 microphone arrays. Continuous audio recording is divided into 10 s audio clips. These audio clips are saved as separate files. Table 3 presents the detailed information of experimental dataset which contains 9 daily domestic activities. All audio clips listed in Table 3 are adopted for domestic activity clustering.

Table 3. Detailed information of experimental dataset.

| Domestic activities | Clips No. |
|---|---|
| Absence (nobody presents in the room) | 18860 |
| Cooking | 5124 |
| Dishwashing | 1424 |
| Eating | 2308 |
| Others (present without doing any relevant activity) | 2060 |
| Social activity (visit, phone call) | 4944 |
| Vacuum cleaning | 972 |
| Watching TV | 18648 |
| Working (typing, mouse click, etc.) | 18644 |
| **Total** | **72984** |

### 3.2. Experimental setup

Both clustering accuracy (CA) and normalized mutual information (NMI) are used as performance metrics. They have been popularly adopted as metrics for clustering, and their details are referred to [31]. The higher their values are, the better the performance is. Main parameters of the proposed method are optimally regulated on experimental dataset and their settings are presented in Table 4.

Table 4. Main parameters settings of the proposed method.

| Type | Parameters settings |
|---|---|
| MFCC | Frame length/overlapping: 40ms/20ms |
| | Dimension of MFCC: 28 |
| CCAN | Pretraining iterations: 100 |
| | Maximum iterations: 2000 |
| | Batch size: 32 |
| | Learning rate: 0.001 |
| | Number of convolution layers (CL): 5 |
| | Number of deconvolution layers (DL): 3 |
| | Size of a convolution kernel: 3×3 |
| | Channel number of CL: [64 64 64 128 128] |
| | Channel number of DL: [128 64 32] |
| | Dimension of deep representation: 9 |
| | Threshold $\varepsilon$: 0.05 |
| | Coefficient $\alpha$: 0.1 |
| | Neuron number of full-connection layer: 1152 |
| | Size of a primary capsule: 6×6×9×16 |
| | Size of a class capsule: 16×9 |
| | Number of primary capsules per layer: 16 |
| | Number of class capsules per layer: 9 |
| | Number of cluster centers: 9 |
| | Value of $max\_iterations$: 3 |

### 3.3. Results and discussions

The proposed method is compared with 5 methods, including: HGTR based method [14], Stacked Autoencoders (SA) based method [32], Capsule Autoencoder Network (CaAN) based method [33], Convolutional Autoencoder Network (CoAN) based method [24], and Long Short-Term Memory (LSTM) based method [34]. The deep embeddings are learned by: SA in [32], CaAN in [33], CoAN in [24], and LSTM in [34]. In contrast, the HGTR in [14] is a shallow hand-crafted feature. The parameters of these methods are set according to the suggestions in respective references and optimally tuned on experimental data. The results obtained by all methods are given in Table 5.

Table 5. Performance comparison of different methods for domestic activities clustering.

| Features | CA (%) | NMI (%) |
|---|---|---|
| HGTR based method [14] | 43.75 | 36.77 |
| SA based method [32] | 45.47 | 38.48 |
| CaAN based method [33] | 54.81 | 46.03 |
| CoAN based method [24] | 49.62 | 42.60 |
| LSTM based method [34] | 50.36 | 43.08 |
| **Proposed method** | **61.91** | **53.84** |

The proposed method achieves CA of 61.91% and obtains absolute gains of 18.16%, 16.44%, 7.10%, 12.29%, and 11.55% over HGTR based method, SA based method, CaAN based method, CoAN based method, and LSTM based method, respectively. In terms of NMI, the proposed method obtains 53.84%, and gets absolute improvements of 17.07%, 15.36%, 7.81%, 11.24%, and 10.76% over HGTR based method, SA based method, CaAN based method, CoAN based method, and LSTM based method, respectively. Based on the results in Table 5, it can be concluded that the proposed method exceeds the state-of-the-art methods in terms of both CA and NMI. In addition, the proposed framework of the CCAN is an effective combination of both the CaAN and the CoAN, and the CCAN outperforms both CaAN and CoAN in our experiments. This result indicate that the convolutional layers and capsule layers have complementary advantages for representation learning.

To visually show the clustering results obtained by our method, spatial distribution of different clusters is illustrated in Fig. 2 which is generated by a Python library: *matplotlib*. Although most audio clips of the same class are clustered to their respective cluster centers, there are still many confusions among different clusters. For example, audio clips of the *Others* are scattered to other clusters, especially the *Absence*. The reasons for leading to the confusions are probably that: the property differences of these classes are not discriminatively represented and there are overlapping regions in the feature distribution of these classes.

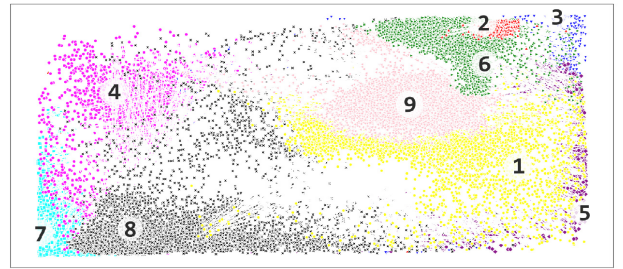

Fig. 2. Visualization of clustering results obtained by the proposed method. Digits in the figure denote different clusters (domestic activities). That is, 1: Absence; 2: Cooking; 3: Dishwashing; 4: Eating; 5: Others; 6: Social activity; 7: Vacuum cleaning; 8: Watching TV; 9: Working.

In summary, our method has been proved to be effective and obtains the most satisfactory results among all methods. However, the domestic activities clustering from audio recordings is still a challenging problem, whose difficulties lie in three aspects. First, it is an unsupervised learning problem without using any labels of audio clips. Second, the data amount of audio clips among different classes are unbalanced. Third, there are overlapping regions in the feature distribution of audio clips among various classes.

### 4. CONCLUSIONS

We tackle a new problem of domestic activity clustering from audio recordings by a CCAN. Our method exceeds the state-of-the-art methods in terms of CA and NMI. Future work includes: discussing the effect of various components of the proposed system, such as capsule network, clustering loss, and reconstruction loss; exploring other networks for deep embedding learning and clustering; and estimating human activities from audio recordings in other situations.